\documentclass[fleqn,usenatbib]{mnras}

\usepackage{newtxtext,newtxmath}
\usepackage[T1]{fontenc}
\usepackage{ae,aecompl}

\usepackage{graphicx}	% Including figure files
\usepackage{amsmath}	% Advanced maths commands
\usepackage{amssymb}	% Extra maths symbols
\usepackage{float}
\usepackage[caption = false]{subfig}
\usepackage[utf8]{inputenc}
\usepackage{ctable}
\usepackage{hyperref}
\usepackage[normalem]{ulem}
\usepackage{chngcntr}
\title[Magnetised winds $\&$ upper planetary atmospheres]{Magnetised winds and their influence in the escaping upper atmosphere of HD 209458b}

\author[Carolina Villarreal D'Angelo et al.]{Carolina Villarreal D'Angelo$^{1}$\thanks{E-mail: csvd@st-andrews.ac.uk}, Alejandro Esquivel$^{2}$, Mat{\'i}as Schneiter$^{3}$,
\newauthor $\&$ Mario Agust{\'i}n Sgr{\'o}$^{3}$
\\
$^{1}$SUPA, School of Physics and Astronomy, University of St Andrews, North Haugh, KY16 9SS, UK\\
$^{2}$Instituto de Ciencias Nucleares, Universidad Nacional Aut\'onoma de M\'exico, AP 70-543, Ciudad de M\'exico 04510, M\'exico\\
$^{3}$ Instituto de Astronom{\'i}a Te{\'orica} y Experimental, Conicet-UNC, Laprida 854, X5000BGR, C{\'o}rdoba, Argentina
}

\date{Accepted 2018 June 6. Received 2018 June 6; in original form 2018 January 2}

\pubyear{2018}

\begin{document}
\label{firstpage}
\pagerange{\pageref{firstpage}--\pageref{lastpage}}
\maketitle

\begin{abstract}
Lyman $\alpha$ observations during an exoplanet transit have proved to be very useful to study the interaction between the stellar wind and the planetary atmosphere. They have been extensively used to constrain planetary system parameters that are not directly observed, such as the planetary mass loss rate.
In this way, Ly~$\alpha$ observations can be a powerful tool to infer the existence of a planetary magnetic field, since it is expected that the latter will affect the escaping planetary material.
To explore the effect that magnetic fields have on the Ly~$\alpha$ absorption of HD 209458b, we run a set of 3D MHD simulations including dipolar magnetic fields for the planet and the star. We assume values for the surface magnetic field at the poles of the planet in the range of $[0$--$5]$~G, and from $1$ to $5$~G at the poles of the star.
Our models also include collisional and photo-ionisation, radiative recombination, and an approximation for the radiation pressure.
Our results show that the magnetic field of the planet and the star change the shape of the Ly~$\alpha$ absorption profile, since it controls the extent of the planetary magnetosphere and the amount of neutral material inside it.
The model that best reproduces the absorption observed in HD~209458b (with canonical values for the stellar wind parameters) corresponds to a dipole planetary field of $\lesssim 1$ gauss at the poles.
\end{abstract}


%%%%%%%%%%%%%%%%%%%%%%%%%%%%%%%%%%%%%%%%%%%%%%%%%%

%%%%%%%%%%%%%%%%% BODY OF PAPER %%%%%%%%%%%%%%%%%%

\section{Introduction}

The past two decades have been revolutionary when it comes to the study of planetary science. Since the confirmation of the first pulsar planet \citep{wolszczan1994}, and the first discovery through the transiting technique \citep{charbonneau2000}, more than two thousand planets beyond our solar system have been detected. Until then, planetary atmospheres were thought to be  subject to a relatively calm environment. Yet, more than two-thirds of the firstly discovered exoplanets were observed so close to their host star ($a \le  0.5$ AU), as to imply a much more active interaction with the stellar wind and radiation.

For some of these exoplanetary systems, observations in the Lyman $\alpha$ line provided evidence of neutral atmospheric material loss, as a consequence of the heating produced by the enormous amount of UV flux received from their host star, e.g. HD 209458b \citep{vidal2003}, HD 189733b \citep{lecavelier2010,lecavelier2012, Jensen2012, benjaffel2013}, Wasp 12b \citep{Fossati2010,Haswell2012, Nichols2015}, GJ 436b \citep{kulow2014,Ehrenreich2015} and possibly 55 Cnc b \citep{Ehrenreich2012}.

Of particular interest is the work of \cite{vidal2003} where the first Ly~$\alpha$ transit observations were analysed, and the presence of an escaping neutral atmosphere was proposed.
The authors found a $10\%$ absorption at {$-100$~km~s$^{-1}$}, an asymmetric line profile with more absorption in the blue than in the red part of the line, and a total absorption of $[15 \pm~4]\,\%$ in the range of {$\pm300$~km~s$^{-1}$}.

Later, transit absorption observations of atomic lines from \ion{H}{i}, \ion{O}{i}, \ion{C}{ii}, \ion{Si}{iii} and \ion{Mg}{i} \citep{vidal2003, vidal2004, ballester2007,Ehrenreich2008,Linsky2010, Jensen2012,vidal-madjar2013} helped to confirm the presence of such a hydrodynamic planetary wind.

We know now that the features present in the Ly~$\alpha$ line during the transit of HD 209458b can be explained by a combination of several physical processes such as the stellar-planetary wind interaction, the stellar radiation pressure, and the charge exchange of planetary neutral atoms with stellar ions, etc.

Numerical and theoretical studies have been carried out with the inclusion of one or several of these processes.
For instance, the expansion of the planetary atmosphere and the resulting
planetary wind, as a consequence of the incident UV stellar flux, have been studied in the works of \cite{murray-clay2009,Koskinen2010,koskinen2013b,guo2011,Salz2016}, among others.
These works give a good description of the lower layers of planetary atmospheres, however, they do not include the dynamical interaction with the stellar wind.

The stellar-planetary wind interaction is considered using hydrodynamic simulations in  \citet{schneiter2007,villarreal2014,schneiter2016} and through particle simulation including the charge exchange process in \citet{erkaev2007,holmstrom2008,tremblin2013,bourrier2013,kislyakova2014} and \citet{christie2016}.
These works were able to partially reproduce the observed absorption in the Ly~$\alpha$ line even though they did not take into account the stellar nor the planetary magnetic fields.

Planetary magnetic fields can shield the atmosphere of the planet from the direct interaction with the stellar wind, as they deflect the stellar winds and prevent their penetration into the lower layers of planetary atmospheres. Several studies indicate that, if present, an exoplanetary magnetic field will ultimately determine the amount of atmospheric material that is lost from the planet \citep{adams2011,Trammell2011,trammell2014,owen2014,Khodachenko2015}. However, the presence of a planetary magnetic field is still an open question.

From theoretical calculations, the expected magnetic moment for hot-Jupiter like planets should be only a few times lower than the magnetic moment of Jupiter ({$\mathcal{M}_\mathrm{J}=1.56\times10^{27}$~A~m$^2$}) \citep{Sanchez-Lavega2004,Durand2009}. Nevertheless, \citet{Christensen2009} predict magnetic field strengths, depending on the internal heat flux, that could be an order of magnitude higher than that of Jupiter. Recently, \citet{Rogers2017} proposed that a small planetary dynamo, due to conductivity variations arising from the strong asymmetric stellar heating, could exist in these type of planets.

The magnetic field value of HD 209458b remains unknown. Moreover, the magnetic field strength for its host star has not yet been detected via spectropolarimetric observations \citep{Mengel2017}.
Analytical calculations for the planetary magnetic moment have given values in the range $[1$--$8]\,\mathcal{M}_\mathrm{J}$ \citep{Sanchez-Lavega2004,Durand2009}.
\cite{kislyakova2014} proposed a magnetic moment of $0.1\,\mathcal{M}_\mathrm{J}$ based on the transit observations in the Ly~$\alpha$ line.
\cite{trammell2014} also employed the Ly~$\alpha$ observations to constrain the magnetic field of HD 209458b. In that work, the authors found that atmospheric material can be supported within closed magnetic field lines in what is known as a wind `dead zone' (also found by \cite{Trammell2011} and \cite{Khodachenko2015}).
When compared to the Ly~$\alpha$ transit absorption, this trapped material can reproduce the observations of \cite{benjaffel2008}  for a planetary magnetic field of $10$~G ($2.8\,\mathcal{M}_J$) or less.
Others works that aimed to explore the effects that the magnetic field has on the atmospheric expansion have adopted similar values \citep[see][]{GrieBsmeier2004, Khodachenko2012,Khodachenko2015,owen2014}. \citet{Erkaev2017} used a 3D MHD model to reproduce the interaction of a non-magnetised planet and the stellar wind. In order to reproduce the Ly~$\alpha$ observations as in \cite{kislyakova2014}, they concluded that the planet should have an intrinsic magnetic moment of about {$[0.13$--$0.22]\,\mathcal{M}_\mathrm{J}$}.

In previous works \citep{schneiter2007,villarreal2014,schneiter2016} we have studied how the hydrodynamic wind-wind interaction shapes the Ly~$\alpha$ profile during the planetary transit. Using 3D hydrodynamics simulations, we correlated the planetary mass loss rate ($\dot M_\mathrm{p}$) with the observed total absorption in this line. \cite{schneiter2007} assumed solar wind conditions for a fully ionised stellar wind and a neutral planetary wind, proposing an upper limit for $\dot M_\mathrm{p}$. \cite{villarreal2014} explored the influence of different stellar and planetary wind conditions and structures for the planetary wind and finally, \cite{schneiter2016} included explicitly the photoionization process in the calculations, thus making the proposed model closer to self-consistent. When contrasted with the Ly~$\alpha$ transit observations, all of them successfully reproduced the main features mentioned in \cite{vidal2003}, but they fail to reproduce the absorption in the red part of the line profile, which we have successfully reproduced in the present work by including magnetic fields.

Until now, there are only a few 3D MHD simulations that consistently model the stellar-planetary wind interaction \citep[see][]{cohen2011a,matsakos2015,Tilley2016} but none of them is based on HD 209458b. This work is an effort to explore how the magnetic field affects the Ly~$\alpha$ line during transit, considering both the star and the planetary magnetic fields with different magnetic moments.
Due to the collisional coupling between ions and neutrals in the upper planetary atmosphere, we expect a measurable influence of the magnetic field on the observed Ly~$\alpha$ profile, as suggested by \cite{trammell2014} and  \cite{Erkaev2017}.
Our 3D MHD models have the star and the planet in the same physical domain. They include a proper treatment of the stellar photoionization and the effects of cooling and heating of the neutral material.

Section \ref{code} introduces the new MHD version of the {\sc guacho} code. Appendix \ref{appendix} presents some of the 1D/2D/3D MHD tests that were performed in order to validate the MHD implementation. The numerical setup is presented in \S\ref{setup}. Results, analysis and comparison with observations are treated in section \S\ref{results}. Finally, the conclusions are given in section \S\ref{conclusions}.

%--------------------------------------------------------
\section{The MHD code: GUACHO}\label{code}

The present work introduces a new version of the 3D hydrodynamic/radiative code {\sc guacho} \citep{esquivel2009,esquivel2013} that solves the ideal magneto-hydrodynamics equations in a Cartesian grid, together with the radiative transfer process in the following form:

\begin{equation}
\frac{\partial \rho}{\partial t} + \nabla \cdot (\rho {\bf u})=0,
\label{eq:cont}
\end{equation}
\begin{equation}
\frac{\partial (\rho {\bf u})}{\partial t} + \nabla \cdot \left[\rho {\bf
  uu}+{\bf I}(P+\frac{ B^2}{8 \pi}) - \frac{{\bf BB}}{4\pi}\right]={\bf f}_\mathrm{g},
\label{eq:mom}
\end{equation}
\begin{equation}
\frac{\partial E}{\partial t} + \nabla \cdot \left[{\bf
    u}(E+P+\frac{ B^2}{4\pi})-\frac{({\bf u \cdot B}) {\bf B}}{4\pi} \right]=G_\mathrm{rad}-L_\mathrm{rad}+ \mathbf{f}_\mathrm{g} \cdot
     {\bf u},
\label{eq:ener}
\end{equation}
\begin{equation}
\frac{\partial {\bf B}}{\partial t} - \mathbf{\nabla \times \left( u \times B\right)}=0,
\label{eq:ind}
\end{equation}
where $\rho$, {\bf u}, P, {\bf B}, and $E$ are respectively the mass density, velocity,
thermal pressure, magnetic field, and (total) energy density. {\bf I} is the identity matrix, $\mathbf{f}_\mathrm{g}$ is the gravitational force, while $G_\mathrm{rad}$ and $L_\mathrm{rad}$ are the
gains and losses due to radiation.
The energy density and thermal pressure are related through an ideal
gas equation of state, that is $E = \rho u^2/2 + P/(\gamma-1)+B^2/8\pi$,
where $\gamma$ is the ratio between specific heat capacities. In our models we have set $\gamma = 1.05$ to reproduce the acceleration of the stellar and planetary wind (as in \cite{matsakos2015,Vidotto2009,Trammell2011}).

The set of Equations \ref{eq:cont}--\ref{eq:ind} is advanced in time
with a second order Godunov method. The fluxes are calculated with the HLLD Riemmann solver of \cite{miyoshi-kusano}, and a linear reconstruction of the primitive variables is applied using the $\mathrm{minmod}$ slope limiter to ensure stability.

As in every MHD code, the magnetic field divergence must be maintained close to zero. For this purpose, the \textit{flux-interpolated} Central Difference scheme developed by \citet{toth2000} was implemented.

The right-hand sides of eqs. \ref{eq:mom} and \ref{eq:ener} show the source terms included in our simulations. We will explain them separately in the following subsections.

\subsection{Gravity and the stellar radiation pressure}
We modelled the gravity of the planet and the star as if the total mass of each object were concentrated at their centre. To include the stellar radiation pressure we use the same approach of previous works \citep{vidal2003,vidal2004,lecavelier2010, villarreal2014,schneiter2016}, and assume that as a first approximation the radiation pressure force can be considered a reduction of the stellar gravity. Thus, the planetary wind feels an effective gravity given by
\begin{equation}
\mathbf{f}_\mathrm{g} = \rho \left[{\bf g_\mathrm{p}}+(1-\mu){\bf g_\star}\right],
\end{equation}
where $\mu$ is the ratio between the radiation pressure and the gravitational force from the star and, $\mathbf{g}_\mathrm{p}$ and $\mathbf{g}_\mathrm{\star}$ are the acceleration due to the gravitational forces from the planet and the star, respectively.
In our models, and assuming that the total Ly~$\alpha$ flux of HD 209458 is the same as that of our Sun at solar minimum ({$3\times10^{11}$~ photons cm$^{-2}$~s$^{-1}$} \citep{vidal1975,Tobiska1997}), we get $\mu=0.7$
\citep{Lemaire2002,Bzowski2008}.

In the current models, we ignore the dependence of $\mu$ with the relative velocities of the H atoms with respect to the star. A more careful treatment of the stellar radiation pressure force that includes such dependency is being currently in consideration (Esquivel et al. in prep), but the preliminary results shows that it is not very relevant for a stellar wind with the characteristics of HD~209458.

\subsection{Heating, cooling and radiative transfer}

The radiative gains and losses are produced by the cooling and heating of the neutral material in the planetary wind. In our models, we account for the processes of photo-ionisation, collisional ionisation and radiative recombination of neutral hydrogen, integrating an additional equation that follows the change of ionisation state of hydrogen together with the gas-dynamic equations:
\begin{equation}
 \begin{split}
  \frac{\partial n_\mathrm{H_I}}{\partial t} + \nabla.(n_\mathrm{H_I}
  {\bf{u}})=  & n_\mathrm{e}\,n_\mathrm{H_{II}}\,\alpha(T)\\
& -n_\mathrm{e}\,n_\mathrm{H_I}\,c(T)-n_\mathrm{H_I}\phi,
 \end{split}
\label{eq:hrate}
\end{equation}
where $\alpha (T)$ is the recombination coefficient, $c(T)$ is the collisional ionisation
coefficient of hydrogen, and $\phi$ the photo-ionisation rate.
For Equation (\ref{eq:hrate}) we assume that all the free electrons come from ionisation of hydrogen, that is $n_\mathrm{e}=n_\mathrm{H_{II}}=(n_\mathrm{H}-n_\mathrm{H_I})$, where $n_
\mathrm{H}=\rho/m_\mathrm{H}$ is the total hydrogen density ($m_\mathrm{H}=1.66\times10^{-24}\,\mathrm{g}$ being the hydrogen mass).

The radiative field of the star is included with the ray tracing method used in \cite{schneiter2016}. The total stellar EUV flux is divided in $10^6$ photon packages that are launched from the stellar surface at random positions in random directions. As these packages travel through the grid, they are attenuated by a factor $\mathrm{e}^{-\Delta\tau}$, with {$\Delta\tau=a_0\,n_\mathrm{{H_I}}\,\Delta l$}. This factor depends on the neutral material within each cell, the path-length ($\Delta_\mathrm{l}$) and the photo-ionisation cross-section that we have assumed to be {$a_0=6.3\times 10^{-18}$~cm$^2$}, considering that all the photons are at the Lyman limit. The photo-ionisation rate $\phi$ is then calculated by equating the ionising photon-rate $S_\star$ with the ionisation per unit time within each cell ($S_\star=n_{\mathrm{H_I}}\,\phi\,\mathrm{dV}$). The contribution of the photo-ionisation rate is then added to Eq. \ref{eq:hrate} and also used to calculate the heating term in Eq. \ref{eq:ener}. The heating per unit time and volume is $\psi=\phi\,E_0$, with an energy gain per ionisation of $E_0=0.86$~eV.

The cooling is computed using the prescription described in \citet{1995MNRAS.275..557B}, which includes contributions from recombination and collisional ionisation of hydrogen, collisional excitation of  the H Ly~$\alpha$ line and [\ion{O}{i}], [\ion{O}{ii}] forbidden lines. The latter are obtained assuming that the excitation is in the low-density regime (for which we solve the 5 level atom problems with the atomic parameters from \citealt{2011aas..book.....P}), and assuming that the ionisation state of oxygen follows that of hydrogen (which can be justified by the efficient charge exchange between them). The oxygen cooling rate is multiplied by a factor of $\sim7$ in order to account for other important coolants (such as C, N and S, see \citealt{1995MNRAS.275..557B}).
At temperatures above $5\times 10^4~\mathrm{K}$ (where oxygen is expected to be more than singly ionised) we switch to a coronal equilibrium cooling curve (in fact the parametrisation of the cooling calculated by \citealt{1976ApJ...204..290R} given in \citealt{1989ApJ...344..404R}).

\section{The numerical setup}\label{setup}

Our models are based on the planetary system HD 209458b, composed by a solar like star and a hot Jupiter like planet. The whole system is simulated in a Cartesian mesh of $460\times230\times460$ cells with a resolution of $3\times10^{-4}$ AU.

All the models have the same geometrical configuration, with a non-rotating star located at the centre of the coordinate system, and a non-rotating planet in a circular orbit with a radius $0.047~\mathrm{AU}$ in the $xz$-plane.

To initialise our models, we fill the entire grid (except inside the planet wind boundary) with the stellar wind using the isothermal Parker wind solution \citep{Parker1958} corresponding to a coronal temperature of $T_\star=1.5 \times 10^6$ K and the dipolar magnetic field of the star. In this medium we overwrite a region around the planet position with the planetary wind structure (see below).
For subsequent times, the stellar and planetary wind configurations are re-imposed at every time-step within spheres of radii $1\,R_\star$ for the stellar wind and $3\,R_\mathrm{p}$ for the planetary wind.
The position of the planetary boundary changes according to the orbit. We also include the orbital speed to the velocity field in the planet wind. All external boundaries of the simulation are treated with outflow boundary conditions.

The simulations evolve until they reach a steady state, which is usually after the planet completes $\sim3/4$ of its orbit.

The physical parameters of the star/planet system \citep{Torres2008}, together with the adopted values for initialising the winds, are presented in Table \ref{tab:1}.

\subsection{The stellar wind}
The initial conditions set inside the star will give rise to a continuously blowing wind. Using the stellar parameters for the solar like star HD 209458, we calculate the values of density, velocity and temperature at the surface of the star ($R_\star$) and extrapolate them as constant values inside the stellar radius to fill the entire stellar volume. Since the values inside this boundary are reimposed at every time-step the flow within this region does not evolve with time.

The velocity at the stellar surface is calculated by means of an isothermal wind model \citep{Parker1958} for a coronal temperature $T_\star=1.5 \times 10^6$ K, using the approximation given in eq. 323 in \cite{lamercassinelli1999}:
\begin{equation}
v_\star= a\left(\frac{v_\mathrm{e}}{2a}\right)^2 \exp\left(\frac{-v_\mathrm{e}}{2a^2} + \frac{2}{3}\right),
\end{equation}
where $a=\sqrt{k_B T/\overline{\mu} m_H}$ is the sound speed, $v_\mathrm{e}=\sqrt{2GM_{\star}/R_{\star}}$ is the escape velocity, and $\overline{\mu} = 0.5$ is the mean molecular weight, assuming a fully ionised H wind.
The value chosen for the coronal temperature in our simulations is in agreement with the one calculated using the formula derived in the work of \cite{johnstone2015} ($\overline{ T}_{cor}= 0.11F_x^{0.26}$) and the value of the X-ray surface flux $F_x$ derived in the work of \cite{sanz-forcada2011}.

Fixing the stellar mass loss rate to the solar value of $2.0\times10^{-14}$ $M_{\odot}$ yr$^{-1}$, we can derive the value of the density at the base of the wind ($\dot{M}_{\star}=4 \pi\rho_{\star}R_{\star}^2v_\star$).

For the stellar magnetic field we assume a dipolar configuration oriented in the $y$-direction (perpendicular to the orbital plane). This is justified by the fact that, for a solar type star, the major contribution to the magnetic field at the position of the planet will be the dipolar component, which in Cartesian coordinates can be written as
\begin{equation}
\mathbf{B}(x,y,z)=\frac{B_{\star}}{2}\left(\frac{R_\star}{r}\right)^3\frac{1}{r^2}\left(3xy,3y^2-r^2,3zy\right),
\end{equation}
where $r^2=x^2+y^2+z^2$ and $B_{\star}$ is the magnitude of the surface stellar magnetic field at the poles. We choose to explore two different values for $B_{\star}$, 1 and 5 gauss, which are typical values for the large-scale solar magnetic field during a solar cycle. Inside the star (where the MHD equations are not evolved), we change the radial dependence of {\bf B} to avoid high magnetic field values at the centre, saturating the value of the magnetic field to that at $R_{\star}/2$, see \citealt{matsakos2015}.

%------------------------- 3d plot---------------

\begin{figure*}
\includegraphics[width = 0.8\textwidth]{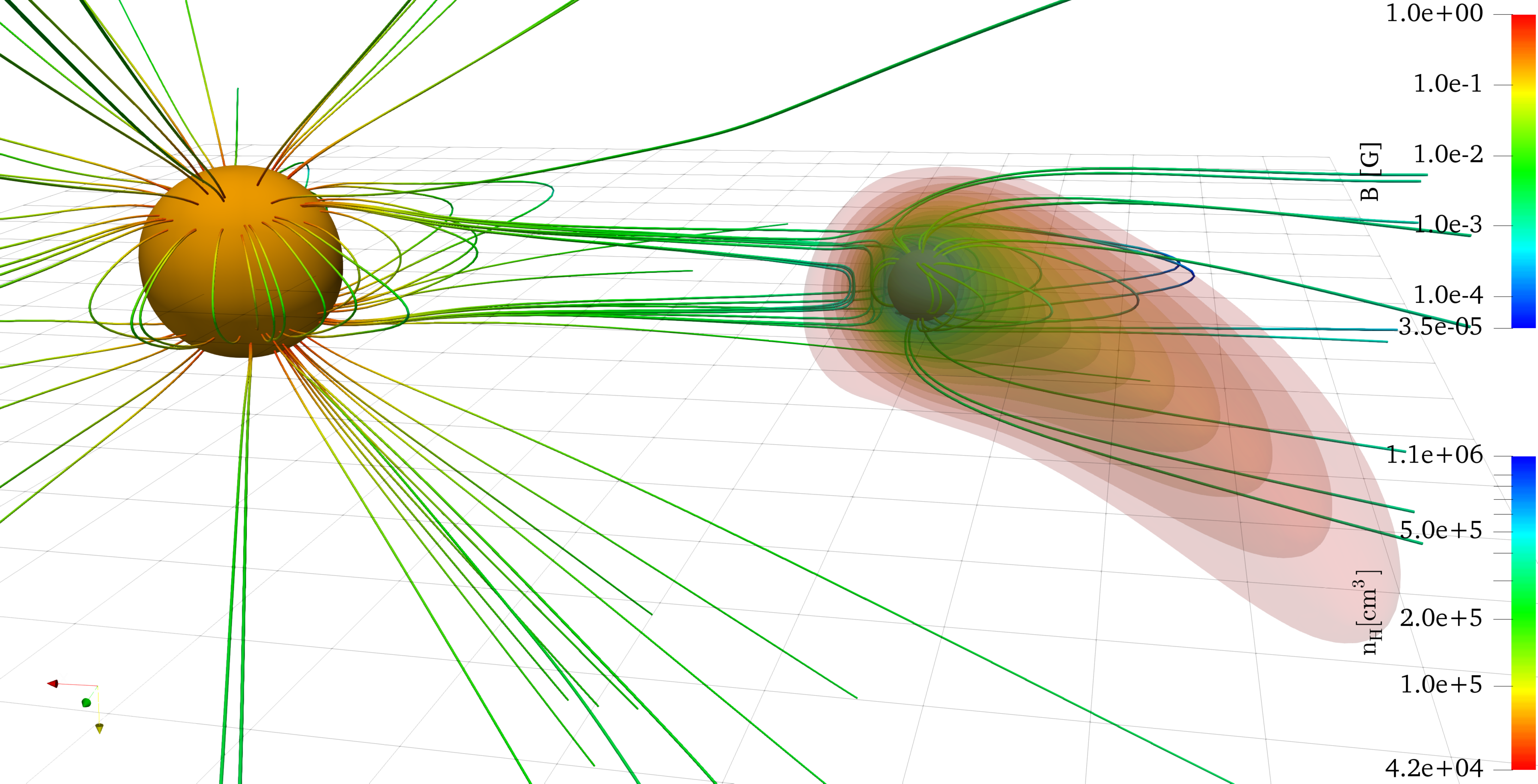}
\caption{3D view of the magnetic field lines coloured by the intensity of the magnetic field for model B1.1. The contours around the planet position show the number density.}
\label{fig:exo}
\end{figure*}
%----------------------------------------------------

The radiative field of the star is simulated with the emission of photons at random positions from the stellar surface. The total amount of photons emitted in random directions is calculated using the stellar luminosity in the EUV, $\log_{10}(L_{EUV}) < 27.74$~erg~s$^{-1}$, derived in the work of \cite{sanz-forcada2011}, which is then divided in $10^6$ photon-packages. The adopted stellar photon rate is therefore ${S_{0}=2.5\times10^{38}\,\mathrm{s^{-1}}}$, which, assuming that each photon is at the Lyman limit, corresponds to a flux of ${F_0=884\,\mathrm{erg\,cm^{-2}\,s^{-1}}}$ at the orbital distance of HD 209458b.

\ctable[
	caption= {Stellar and planetary winds parameters employed in the simulations for the system HD 2019458},
    label= {tab:1}
]{llc}{
}{
	\toprule
    \toprule
    {\bf Stellar parameters}                 & Sym.                 & HD 2019458\\
    \midrule
    Radius $[R_\odot]$                       & $R_{\star}$          & $1.2$\\
    Mass $[M_\odot]$                         & $M_{\star}$          & $1.1$\\
    Wind temperature [MK]                    & $T_\mathrm{\star}$   & $1.5$\\
    Mass loss rate $[M_{\odot}~\mathrm{yr}^{-1}]$     & $\dot{M_{\star}}$    & $2.0\times10^{-14}$\\
    Photon rate $[$s$^{-1}]$                   & $S_0$                & $2.5\times 10^{38}$\\
    \midrule
    \midrule
    {\bf Planetary  parameters}              & Sym.                 & HD 209458b\\
    \midrule
    Radius $[R_\mathrm{J}]$                         & $R_\mathrm{p}$       & $1.38$\\
    Mass $[M_\mathrm{J}]$                           & $M_\mathrm{p}$       & $0.67$\\
    Orbital period [d]                       & $\tau_\mathrm{p}$       & $3.52$\\
    Inclination [{\degr}]                    & $i$                  & $86.71$\\
    Wind launch radius $[R_\mathrm{p}]$               & $R_\mathrm{w,p}$     & $3$\\
    Wind velocity at $R_\mathrm{w,p}$ [km s$^{-1}$] & $v_\mathrm{p}$       & $10$\\
    Wind temperature at $R_\mathrm{w,p}$ [K]        & $T_\mathrm{p}$       & $1 \times10^4$\\
    Ionisation fraction at $R_\mathrm{w,p}$         & $\chi_\mathrm{p}$                & 0.8\\
    Mass loss rate [g s$^{-1}]$              & $\dot{M}_\mathrm{p}$ & $2 \times 10^{10}$\\
    \bottomrule
    \bottomrule
}

\subsection{The planetary wind}

In simulating the wind of HD 209458b we are making the assumption that the upper part of the planetary atmosphere is under an hydrodynamic escape as discussed in \citet{murray-clay2009,guo2011,Salz2016}. The approach is the same as the one implemented on the star, specifying the initial values of the hydro-dynamical variables for the planetary wind inside a radius, which in this case is $3\,R_\mathrm{p}$. At smaller radii the values are kept constant. Launching the planetary  wind at this distance allows us to have a lower resolution than the one that would be required to model the generation of the wind from its base ($1\,R_\mathrm{p}$).
The adopted wind parameters at $3\,R_\mathrm{p}$ for an $80\%$ ionised hydrogen atmosphere are presented in Table \ref{tab:1} and are in agreement with the standard planetary wind model derived in \cite{murray-clay2009} for the case of an energy limited driven wind. A velocity of $10$~km~s$^{-1}$ and a temperature of $10^4$~K are then imposed every time step at this boundary.

The planetary density is obtained from its mass loss rate, taken to be $\dot M_\mathrm{p}=2\times10^{10}$~g~s$^{-1}$. This value is consistent with our previous works \citep{villarreal2014,schneiter2016} and it is also in agreement with recent estimations derived from observations \citep{salz2015}.

The planet is considered to support a dipole magnetic field, with the same orientation as the stellar magnetic field (i.e. perpendicular to the orbital plane).
\begin{equation}
\mathbf{B}(x',y',z')=\frac{B_\mathrm{p}}{2}\left(\frac{R_\mathrm{p}}{r'}\right)^3\frac{1}{r'^2}\left(3x'y',3y'^2-r'^2,3z'y'\right),
\label{bp}
\end{equation}
where $r'^2=x^2+y'^2+z'^2$ measured from a coordinate system centred at the planet position, and $B_\mathrm{p}$ is the surface magnetic field value at the poles.
Three different values for the polar magnetic field are employed: 0, 1 and 5 gauss, corresponding to a magnetic moment between $[0-1.54] \, \mathcal{M}_\mathrm{J}$.
Since we are imposing the planetary magnetic field from $3\, R_\mathrm{p}$, the magnitude of $B_\mathrm{p}$ in Eq. \ref{bp} is scaled to this radius.
Even though the adopted values are smaller than those of Jupiter's magnetic field, they are useful in giving an upper limit for possible planetary magnetic field values for these types of planets.
As with the star, the radial dependence of the magnetic field for  $r'\le R_\mathrm{p}/2$ is modified and the value of $B_\mathrm{p}$ is set to the corresponding value at this radius, avoiding high magnetic field values at the centre of the planet.

Our initial conditions result in a subsonic planetary wind ($a\sim12.3$ km s$^{-1}$) launched from a region inside the corresponding sonic radius. The planetary wind will then adjust itself to the conditions outside the launching radius.  In all our models, the planetary wind is never entirely suppressed by the stellar wind pressure. However, in the models with large planetary magnetic fields, we see evidence of wind suppression at the equator near the wind launching region. This could correspond to regions where the magnetic pressure dominates \citep{Trammell2011}, but our resolution is insufficient to make any conclusions.

\section{Results \& Discussion}\label{results}

To investigate the role of magnetic fields on the neutral material carried with planetary wind, we run a total of five models. Each model has a different value for the polar magnetic field at the surface of the star ($B_\star$) and the planet ($B_p$); the remaining parameters for the initial conditions for both winds (stellar and planetary) are shown in Table \ref{tab:2}.

A common feature of the interaction between the planetary and stellar wind is the presence of a shock-like region where both winds meet.
The location where the planetary wind effectively stops the stellar wind, the stand-off distance, is key in controlling the extent of the planetary magnetosphere, and it will depend on the pressure balance at this point.
The stand-off distance ($R_0$) for all the models, measured along the line that joins the star and the planet,  is presented in Table \ref{tab:2}.
Trailing the planet, a cometary tail composed of material from the stellar and planetary winds is formed, a known feature of the wind-wind interaction found in previous works \citep{schneiter2007,villarreal2014,schneiter2016}. A 3D  rendering of model B1.1 is shown on Figure \ref{fig:exo}. Magnetic field lines coloured according to the value of the total magnetic field are shown. The planet is surrounded by contours of the number density that shows the extent of the cometary tail.

\begin{figure*}
\includegraphics[width = \textwidth]{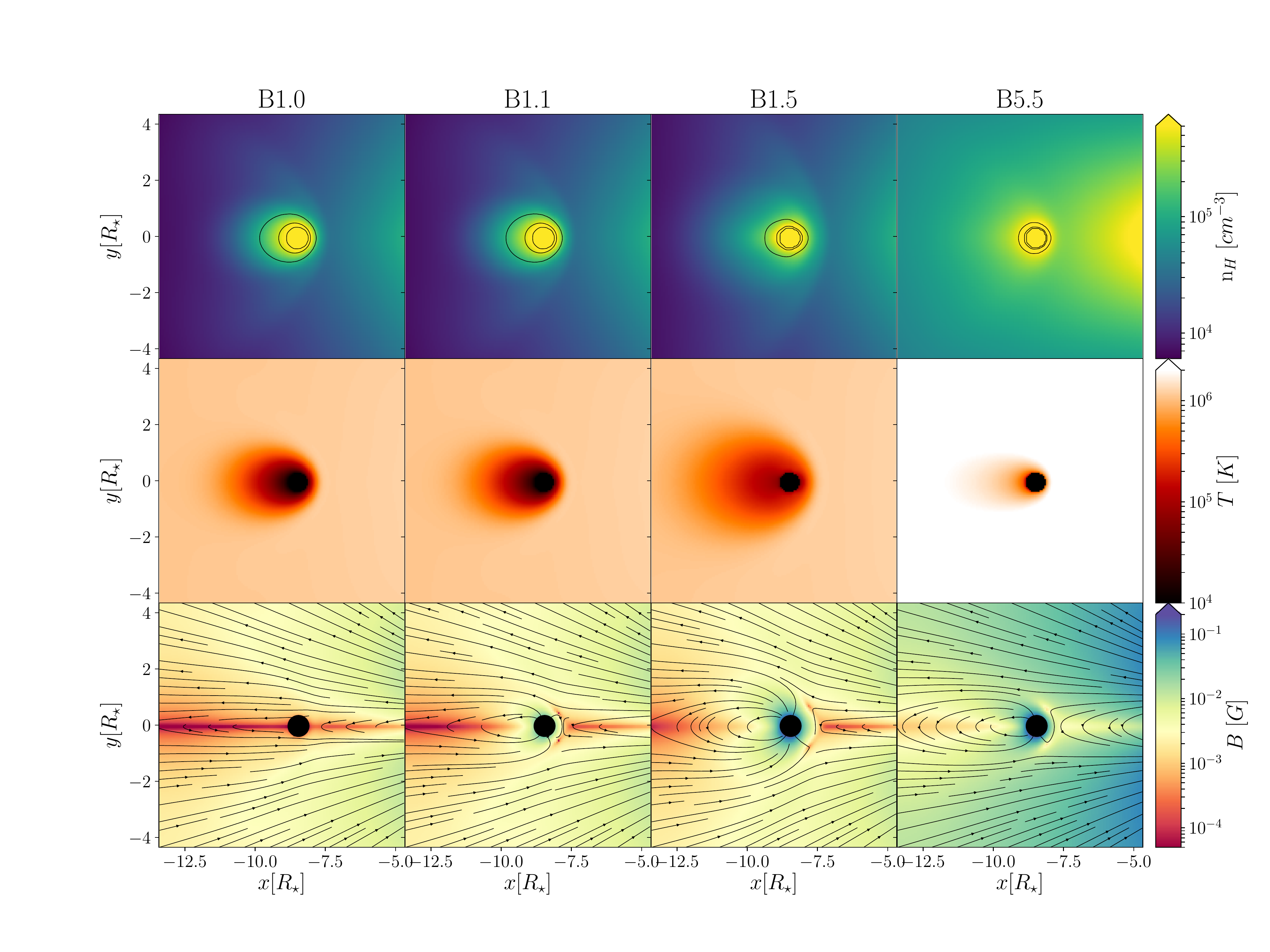}
\caption{Contour plots of density (top row), temperature (middle row) and magnetic field value (bottom row), for a cut in the $xy$-plane (i.e., perpendicular to the orbital plane) centred at the position of planet. In the density plots, the ionisation fraction for values of 0.85, 0.95 and 0.999 is denoted by black contours around the planet.}
\label{fig:rho}
\end{figure*}

To analyse the characteristics of the different models we present in
Figure \ref{fig:rho} the density, temperature and magnetic field values for four of the five models in the $xy$-plane (i.e. perpendicular to the orbital plane) centred at the planet position. In this figure, the star is at the right edge of each panel magenta(also in the $xy$-plane) but out of the field of view. Models B5.1 and B5.5 result in very similar stratification and for this reason we only show model B5.5. This is also true for models B1.0 and B1.1 but, in this case, model B1.0 shows the interaction of the stellar wind with a non-magnetised body.

The first three columns of Figure \ref{fig:rho} correspond to models with $B_{\star}=1$ G characterised by a slower and less dense stellar wind in the vicinity of the planet. In the equatorial plane ($y=0$) a pronounced current sheet develops as a consequence of the stretching of the stellar magnetic field lines, dragged by the stellar wind. The magnetic field strength (bottom row) there becomes almost negligible indicating that the phenomenon of re-connection is taking place. Possible re-connection sites are also visible at the nose of the planetary magnetosphere, at both sides of the equatorial current sheet, for models B1.1 and B1.5. At these points, the planetary magnetic field lines will connect with the magnetic field of the star. In these cases, no Alfvén wings \citep{Neubauer1980} would develop because the stellar wind at the planet position is already super-Alfv\'enic (as well as super-sonic).

The last column in Fig. \ref{fig:rho} shows one of the models with a higher stellar magnetic field value ($B_\star=5$ G). Increasing $B_{\star}$ leads to an increase in the temperature and velocity of the stellar wind \citep{Vidotto2009}. With a higher magnetic tension, the magnetic loops are stretched but not open as in the case of $B_\star=1$~G, remaining closed at the planet position. The  planet, orbiting within a denser and hotter environment, does not suffer from re-connection events between its magnetic field and the stellar one in the direction of the star.

As the stellar wind flows around the planet a cavity forms and is filled with the planet's atmospheric material and magnetic field (when it is present). The properties of the material inside this cavity are governed by the planetary magnetic field, the shock with the stellar wind and the radiative field of the star. This last one will only have a significant influence if the neutral density inside this cavity is optically thick to the stellar photons.

In the case of a non-magnetised planet (model B1.0), the planetary wind fills an region around the planet of lower temperature and higher density than its surrounding. This region's volume and its temperature and density stratification are almost identical to what is observed in model B1.1, where $B_\mathrm{p}=1$ G, suggesting that a planetary magnetic field value less then $1$~G has the same influence in stopping the stellar wind than in the case of a non-magnetised planet. In both cases, a negligible magnetic pressure contribution is present in the planetary side, thus the sum of thermal and ram pressures are responsible for deflecting the stellar wind. When the planetary magnetic field is increased, as in the model B1.5, the resulting cavity has a larger size, in comparison with B1.0 and B1.1 models. The planetary wind gets more easily accelerated along open magnetic field lines (mainly at the poles), and at the same time, the higher magnetic tension at the equator creates a wind `dead-zone' that gets filled with lower temperature material. Both effects result in a larger cavity.

The stand-off distance at the sub-stellar point, presented in Table \ref{tab:2} confirms that models B1.0 and B1.1 share the properties of the resulting cavity since they have the same value of $R_0\sim 7\,R_p$. Similar values have been found by \cite{Weber2017} and \cite{Khodachenko2012} when the contribution of a magneto-disc in the shape of the magnetosphere is taken into account. Smaller values have also been found for HD 209458b in the works of \cite{GrieBsmeier2004} and \cite{kislyakova2014}, although the value of the stellar wind's velocity and density employed in these works were much higher than in our models. A higher stand-off distance is reached for model B1.5, where the magnetic field of the planet is now playing a more important role in the pressure balance.
Model B5.1 and B5.5 present a more compressed planetary magnetosphere, with stand-off distances near $4\,R_\mathrm{p}$. The shorter standoff distance is a result of the increased ram pressure on the stellar side (due to the strong stellar wind). For these cases, the value of the planetary magnetic field has little effect in the properties inside the magnetospheric cavity (the stratification of models B5.1 and 5.5 are remarkably similar). This is due to the fact that the shock with the stellar wind determines the temperature and the velocity of the material of this region.

The ionisation fraction of the planetary wind ($\chi$), is plotted in the first row of Figure \ref{fig:rho}. The three black contour levels around the planet represent the values of $\chi= [0.85, 0.95, 0.999]$. For $B_{\star} =5$~G, a very compressed planetary magnetosphere occurs, and these contours are found very close to the planet, showing that almost all planetary material becomes ionised down to very near the base of the wind.
When the planetary magnetosphere is more expanded, the planetary wind can remain partially neutral (up to $5\,\%$) further from the wind base ($3\,R_\mathrm{p}$), and more so in the tail direction, as can be seen in models with $B_\star=1$ G.
As we mentioned above, a higher value of the planetary magnetic field increase the velocity and temperature in the wind. Hence, ions in the planetary atmosphere are accelerated more and collisional ionisation become more effective. This can be seen in model B1.5, where the material inside the planetary magnetosphere has a temperature higher than $10^5$~K, and neutral material is more concentrated around the planet than in models B1.0 and B1.1. In this case, the shape of the partially neutral region, even when the magnetosphere is larger, is small compared to models with $B_\mathrm{p}=1$~G.

%-----------------------Table 2----------------------------------
\ctable[
	caption={Models characteristics and the estimated values for the stand-off distance at the sub-stellar point. Last column shows the integrated Ly~$\alpha$ absorption in the velocity range of $\pm 300$ km s$^{-1}$.},
    label= {tab:2},
]{lcccc}{
}{
 \toprule
    {\bf Models} & $B_{\star}$ [G]	& $B_\mathrm{p}$ [G]	& $R_0$ [$R_\mathrm{p}$] & (1-I/I$_\star$) [$\%$]\\
    \midrule
    B1.0	 & 1				& 0			& 7             & 12.1 \\
    B1.1	 & 1				& 1			& 7             & 12.1 \\
    B1.5	 & 1				& 5			& 9             & 10.7 \\
    B5.1	 & 5				& 1			& 4             & 12.7 \\
    B5.5	 & 5 			    & 5			& 4             & 13.4 \\
    \bottomrule
}
%------------------------------------------------------------
%------------Figure of contours for 3 models-----------------
%------------------------------------------------------------

\subsection{Lyman $\alpha$ profile}
The transit absorption of the planet in the Ly~$\alpha$ line will be studied in this section, using the same method as in \cite{villarreal2014} and \cite{schneiter2016,schneiter2007}. The optical depth as a function of the velocity along the line of sight (los) is given by:

\begin{equation}
  \tau_\mathrm{v_{los}} = \int n_\mathrm{HI}\, a_0 \, \varphi\left( \Delta v_\mathrm{los} \right) \mathrm{d}s ,
\label{eq:tau}
\end{equation}

%----------------Figure of tau-------------------------------
\begin{figure*}
\includegraphics[width = 0.8\textwidth]{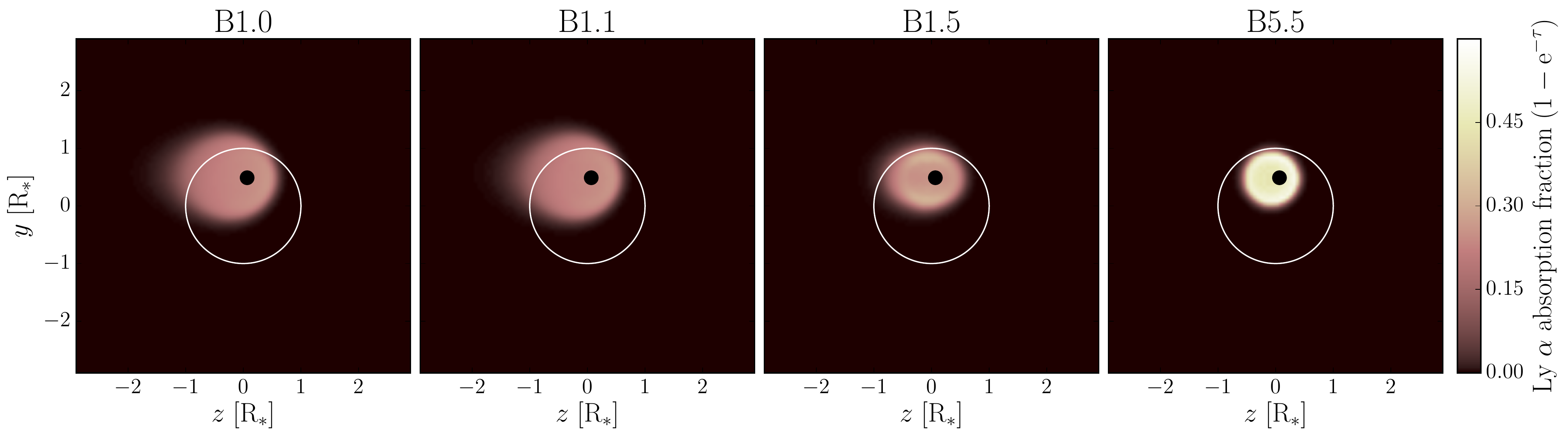}
\caption{Lyman $\alpha$ absorption fraction ($1-e^{-\tau}$), integrated between [-300, 300] km s$^{-1}$, for the four models presented in Fig. \ref{fig:rho}. The plot is a zoom at the position of the star (white circle) at the moment of mid-transit. In order to calculate the absorption in the Ly~$\alpha$ line we have taking into account the orbital inclination of the planet, represented with the black filled circle. }
\label{fig:tau}
\end{figure*}
%------------------------------------------------------------

%------------Figure of Lyman alpha profile-------------------
\begin{figure*}
\subfloat[Normalised Ly~$\alpha$ flux]{\includegraphics[width = 0.5\textwidth]{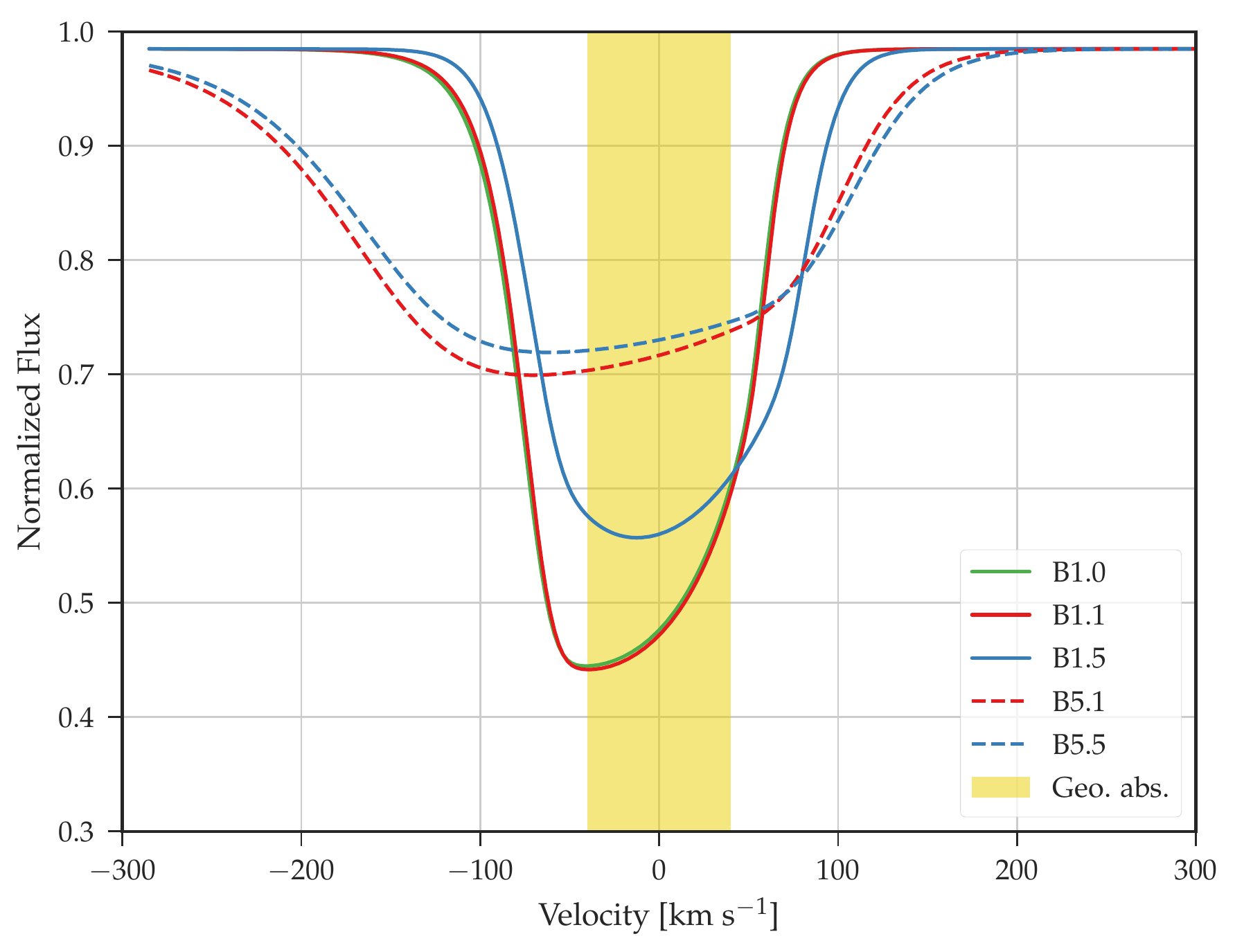}}
\subfloat[Observed Ly~$\alpha$ profile]{\includegraphics[width = 0.5\textwidth]{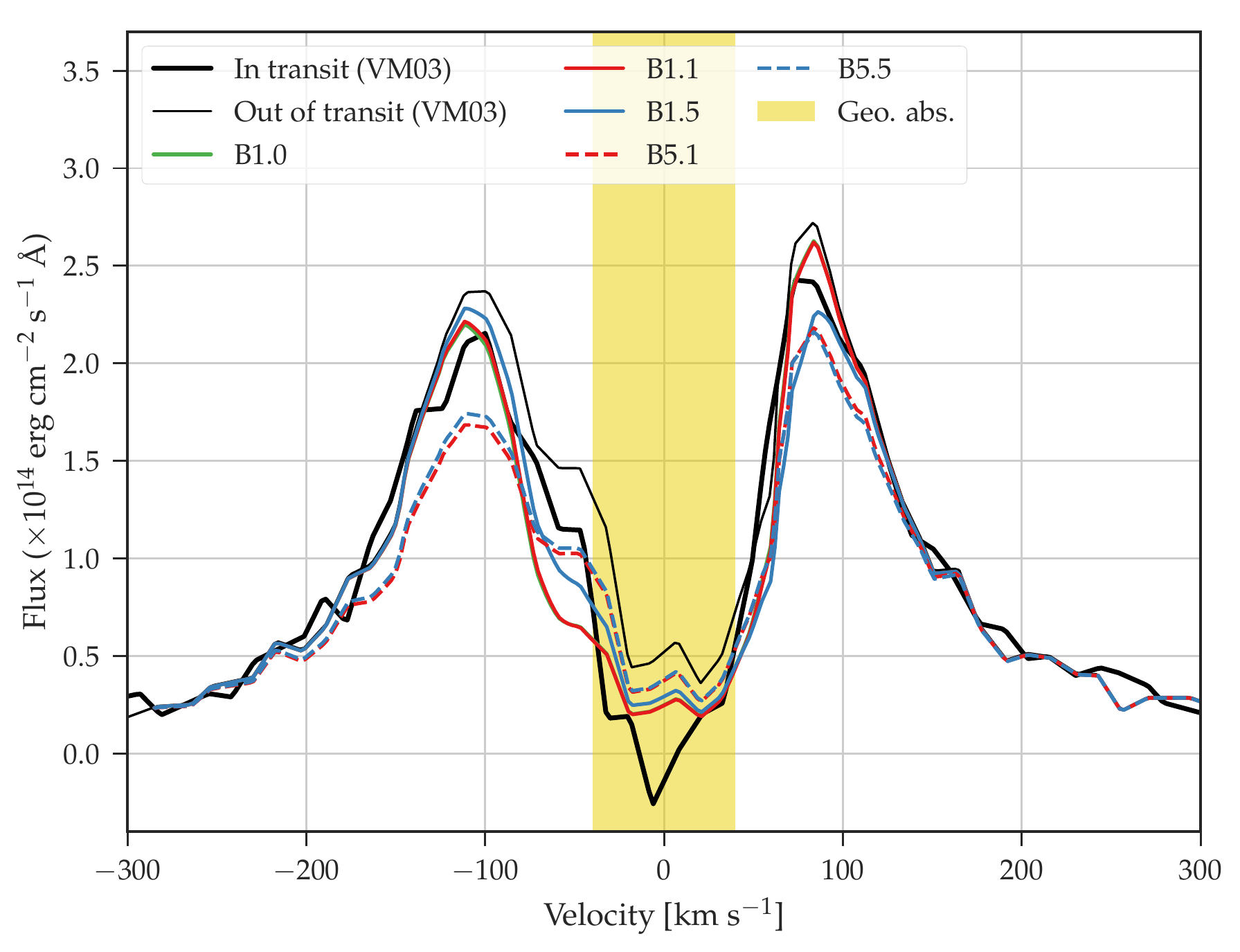}}
\caption{a) Normalised stellar transmission as a function of the line of sight velocity in the Ly~$\alpha$ line averaged over the stellar disk as seen by an observer. b) Comparison with the observational line profile during and off transit, taken from the work of \protect\cite{vidal2003}. The yellow stripe in both graphics correspond to part of the line contaminated with the geo-coronal glow. In both panels, solid lines are used to represent models with $B_{\star}=1$ G and dashed lines are used for models with $B_{\star}=5$ G, while colours represent different values for the planetary magnetic field: green, red and blue show models with $B_p=0,1$ and $5$ G, respectively.}
\label{fig:profile}
\end{figure*}
%------------------------------------------------------------

where $\varphi(\Delta v_\mathrm{los})$ is a Gaussian line-profile\footnote{ We have also computed the absorption profiles using a full Voigt profile and found that the results are virtually equal to the ones obtained with a Gaussian profile. The reasons for this are the low densities ($< 1\times10^{5}$ cm$^{-3}$), and the high temperatures ($>\times 10^5$ K) that the planetary neutral material has close to the planet, responsible for the  absorption found between $[\sim \pm 150]$ km s$^{-1}$.} , $a_0=0.01105$~cm$^2$  is the Ly~$\alpha$ cross-section at the line centre \citep{Osterbrock1989}, $n_\mathrm{HI}$ is the neutral density and $ds$ is the length measured along the line of sight. In our calculations, the velocities are projected along the los-direction ($-x$-direction) and tilted $3.29{\degr}$ around the $z$ axis in order to resemble the orbital inclination ($i=86.71{\degr}$) of the planet as seen from Earth. The integration of $\tau_\mathrm{v_{los}}$ is made from the surface of the star to the end of the computational domain. We let the planet complete almost $5/9$ of its orbit to compute the transit absorption.
The Lyman $\alpha$ absorption fraction $1-e^{-\tau}$ at mid-transit, showed in Fig. \ref{fig:tau}, is then calculated by integrating the optical depth within the velocity range [-300,300] km s$^{-1}$ over 250 channels.

For every velocity channel we have also calculated the normalised stellar transmission during transit integrated over the stellar disc (using the stellar photospheric radius of $1.2\,R_\star$). We do not include the stellar limb-darkening and we assume a constant stellar emission over the adopted velocity range.
%The stellar emission during transit relative to the out of transit emission, $I_v/I_{\star} = \mathrm{e}^{-\tau_\mathrm{v_{los}}}$, is then calculated within the velocity range [-300, 300] km s$^{-1}$ over 250 channels and integrated over the stellar disc (using the stellar photospheric radius of $1.2\,R_\star$). We do not include the stellar limb-darkening and we assume a constant stellar emission over the adopted velocity range.
%
The results are shown in Figure \ref{fig:profile}. Solid lines are used to represent models with $B_{\star}=1$ G and dashed lines are used for models with $B_{\star}=5$ G, while colours represent different values for the planetary magnetic field: green, red and blue show models with $B_\mathrm{p}=0,~1$ and $5$ G respectively.
We compare our results with the observations from \cite{vidal2003} in the same figure. The yellow stripe in both plots represent the part of the Lyman $\alpha$ line that is contaminated with the geo-coronal emission, which is removed from the analysis in order to compare with the observations.

Our calculated line profiles show that models with a higher value of the stellar magnetic field ($B_{\star}=5$ G) present a wider and shallower line profile than models with $B_{\star}=1$ G.
When $B_{\star}$ is $1$ G, the absorption profiles expand a narrow velocity range, from $[-150,120]$ km s$^{-1}$, with a deeper absorption value at the smallest velocities. Models B1.0 and B1.1 have the same absorption profile since they present almost the same physical response to the interaction with the stellar wind. It is for these models that we found $10\,\%$ absorption at $-100$ km s$^{-1}$ and $5\,\%$ absorption at $70$ km s$^{-1}$, in agreement with the observations of \cite{vidal2003,vidal2008}. The distribution of these neutrals, with an average temperature of  $8\times 10^4$ K, is asymmetric with a more extended region oriented towards the observer, as we can see from the line profile. This asymmetric distribution is also visible in Figure \ref{fig:tau}.

When $B_\mathrm{p}=5$ G (model B1.5), the calculated line profile shows absorption at larger velocities towards the star (reaching a $5\,\%$ at $100$ km s$^{-1}$). The shape of the profile is more symmetric when comparing with that obtained for models B1.0 and B1.1 (see also Fig. \ref{fig:tau}), as a result of a more important contribution from the planetary magnetic field. The material that remains neutral inside the magnetospheric cavity reaches temperatures of roughly $2\times 10^5$ K.

For the models with $B_{\star}=5$ G, the planetary wind is embedded inside a hotter environment, and with higher stellar ram pressure. As a result of the shock with the stellar wind, the planetary material gets heated to temperatures of up to $1\times10^6$ K. Nearly $0.1\%$ of the planetary wind
($\sim 2\times10^{-21}$ g cm$^{-3}$) remains neutral under these conditions and acquires velocities around $150$ km s$^{-1}$ in both wings of the profile. Higher velocities are reached towards the blue, driven by the pressure of the stellar wind.
For these models, the different values of the planetary magnetic field that we used make only a marginal difference in the absorption profile as shown in Fig. \ref{fig:profile}.

The absorption profiles presented in Fig. \ref{fig:profile} can be compared with those presented in fig. 5 of \cite{schneiter2016} for models B2b, and C2b. These models have the same initial conditions for launching the planetary wind and comparable stellar wind properties at the position of the planet, but they do not include a magnetic field. Model B2b has stellar wind conditions comparable with models that have $B_\star=1$ G in our work, and model C2b has conditions similar to models with $B_\star=5$~G, except for a lower density of the stellar wind near the planet. The main difference between the hydrodynamic models of \cite{schneiter2016} and our MHD models is that the new computed Ly~$\alpha$ profiles have a higher absorption in the red part of the line. For the models with $B_\star=1~\mathrm{G}$ the profiles of these MHD models have a lower radial velocity (towards the observer), which is due to the fact that the fields lines are not entirely opened away from the star and the material does not flow as freely in this direction.
The absorption profiles of the models with $B_\star=5~\mathrm{G}$ are broader compared with model C2b of \citet{schneiter2016}, but this can be attributed to the difference in the stellar wind ram pressure, given the differences in density of the stellar winds.

The total Ly~$\alpha$ absorption during transit is computed by integrating the line profile over the velocity range $\pm 300$~km~s$^{-1}$, excluding the part of the line contaminated with the geo-coronal emission. The results for all the models are presented in the last column of Table \ref{tab:2}. These values can be compared with the one obtained in the work of \cite{vidal2003} where $[15 \pm 4]\,\%$ of absorption during transit was found. Our results show that the total absorption in the Ly~$\alpha$ line is not strictly correlated with the stand-off distance (the magnetosphere size). For instance,  model B1.5 has a larger stand-off distance but the smallest value for the total absorption in Ly~$\alpha$ during transit.
A more accelerated planetary wind leads to an increase of collisions between ions and neutrals, producing a larger degree of ionisation close to the planet, with more ionisation at the poles (see Fig. \ref{fig:tau}). Models B1.0 and B1.1 have a total absorption during transit of $12\,\%$ that agrees, within the errors, with the value from \cite{vidal2003}.
The highest values ($\sim 13\,\%$) for the integrated line profile are found for models B5.1 and B5.5. When compared with the observations, these models produced a more pronounced absorption during transit as shown in Figure \ref{fig:profile}b. In these cases, the stand-off distance is very close to the planet, indicating a more compressed planetary magnetosphere than in the previous models, due to the stronger stellar winds. This wind also produces a stronger shock, raising the temperature of the material behind up to $\sim10^6~\mathrm{K}$, producing a broad absorption profile from the few surviving neutrals in the wake.

\section{conclusions}\label{conclusions}
Our aim was to study the effects that the presence of dipolar magnetic fields, in the star and the planet, have on the escaping neutral material from the atmosphere of HD 209458b. In doing this, we have assumed that the star could have a dipole magnetic field of magnitude $1$ and $5$~G (at the poles). For the planet we choose to model a non-magnetised planet, and two other examples with a dipole magnetic with magnitudes of $1$ and $5$~G (at the poles). These values correspond to a planetary magnetic moment of $4.8\times 10^{26}$ and $2.4\times 10^{27}$~A~m$^2$ respectively.

We have found that the interaction between both winds leads to the development of a magnetospheric cavity around the planet inside which a partially ionised planetary wind is able to expand. For values of $B_\star=1$ G, this cavity still forms even when the planet magnetic field is set to zero. Moreover, virtually the same region is developed when $B_\mathrm{p}$ is increased from zero to $1$ G, indicating that for lower planetary magnetic field values, the magnetic pressure of the planet is not a major contribution in stopping the stellar wind. These two models also share the same stand-off distance ($\sim 7\,R_\mathrm{p}$) and the same total absorption of $12\,\%$ when we integrated the Lyman $\alpha$ profile in the velocity range of $\pm 300$ km s$^{-1}$ during a transit. When comparing with the observational data, these models are the ones that best reproduce the absorption profile during transit from the work of \cite{vidal2003}. From these observations, the authors also found a total absorption during transit of $[15\pm 4]\,\%$ when integrating the Ly~$\alpha$ line within the same velocity range. For a larger value of the planetary magnetic field, model B1.5 shows that the degree of ionisation of the planetary neutral material is higher as a consequence of the increase in the planetary wind velocity and temperature. Hence, for this model, the total absorption ($\sim 10\,\%$) from the Ly~$\alpha$ line is less than what observations show.

When $B_{\star}$ is higher, a more compressed planetary magnetosphere is formed, and the neutral material piles up close to the planet. Our models show that for these cases, the total absorption in the Ly~$\alpha$ line is about $13\,\%$ during transit, but absorption of $30\,\%$ is found for velocities around $-100$ km s$^{-1}$ and beyond, contradicting the observations.

We confirm that magnetic fields are an important aspect of the stellar and planetary wind interaction. Comparing with our previous work \citep{villarreal2014,schneiter2016}, they help to better reproduce the observed absorption in the red wing of the Lyman $\alpha$ line. The planetary magnetic field can have an influence in stopping the stellar wind, but only for values higher than or equal to $5$~G will it also influence the amount of neutral material present in the planetary wind. It is encouraging to see that for our initial set of parameters chosen the model that best reproduces the Ly $\alpha$ observations of \cite{vidal2003} suggests that the magnetic field of HD 209458b could be less than or equal to $1$ G (in agreement with \citep{kislyakova2014}). We must note that this estimation depends on a large number of model parameters, and a more self-consistent modelling of the generation of the planetary wind is necessary in order to confirm such a conclusion.

\section*{Acknowledgements}
The authors thanks the referee for the comments and suggestions that helped to improve this work.
CVD acknowledge STFC (ST/M001296/1) and the PhD fellowship from CONICET.
MAS acknowledges support from the CONICET via an Assistant Research Fellowship. The authors want to thank to the Centro de Simulaci{\'o}n Computacional para Aplicaciones Tecnol{\'o}gicas - CONICET, where some of the test of the code was run. AE acknowledge support from CONACYT DGAPA-PAPIIT (UNAM) grants IN 109715 and IG 100516.

%%%%%%%%%%%%%%%%%%%%%%%%%%%%%%%%%%%%%%%%%%%%%%%%%%


\bibliographystyle{mnras}
\bibliography{references}
%%%%%%%%%%%%%%%%%%%%%%%%%%%%%%%%%%%%%%%%%%%%%%%%%%

%%%%%%%%%%%%%%%%% APPENDICES %%%%%%%%%%%%%%%%%%%%%

\appendix
\counterwithin{figure}{section}
\section{MHD implementation on GUACHO}
\label{appendix}

We present in this work the new version of the code {\sc guacho}, that includes the equations for solving the ideal MHD problem in a 3D Cartesian grid. The {\sc guacho } code is a free access code and it is maintained in \url{http://github.com/esquivas/guacho}, from where it can be downloaded.

As we state in section \ref{code}, the set of Equations \ref{eq:cont}--\ref{eq:ind} are advanced in time with a second order Godunov scheme.
The code allows the user to choose between two approximate Riemann solvers, the
HLLE solver \citep[][quite robust, but somewhat diffusive]{Einfeldt91}, and the
less diffusive HLLD  of \citet{miyoshi-kusano}.
To achieve second order accuracy, the method performs a linear reconstruction of the primitive variables to solve the Riemann problem at each cell interface. To ensure monotonicity a number of slope limiters for the linear reconstruction step are available, the work presented here makes use of the $\mathrm{minmod}$ (the most robust/diffusive).

In order to keep the divergence of the magnetic field equal to zero, two methods have been implemented in the code: the 8-wave method proposed by \cite{Powell99} and the method of Constrained Transport/Central Difference proposed by \cite{toth2000}. The former considers the MHD equations keeping the terms with
$\nabla \cdot \mathbf{B}$, which is zero theoretically, but slightly non-zero when calculated numerically. Such terms are treated as sources, allowing the spurious magnetic field divergence to behave as a scalar which is transported out of the grid without amplification.
The second method is more elaborated, including an additional step after advancing the equations to correct the magnetic field, and keeps divergence down to machine precision.

To verify that the numerical scheme implemented in {\sc guacho} behaves correctly, we reproduce several 1D, 2D and 3D tests, that are commonly used in the literature to validate MHD codes. All of them can be compared with results from other MHD codes that are widely used, such as {\sc athena}\citep{Stone2008}, or {\sc pluto} \citep{2007ApJS..170..228M}.

\subsection*{Brio-Wu 1D}
This 1D test, proposed by \cite{BrioWu88}, is a MHD version of the Sod shock tube hydro-dynamical problem.
It is a suitable problem to check if MHD codes can accurately resolve shocks, rarefaction, contact discontinuities and other simple MHD structures.

The initial conditions are set at $t = 0$. The entire domain $x \in (-0.5,0.5)$ is filled with a gas with $\gamma = 2$ in two uniform states, separated by a discontinuity at $x=0$. The left side of the domain is initialised to $(\rho,v_x,v_y,v_z,B_x,B_y,B_z,p) = (1,0,0,0.75,1,0,1)$ while in the right side, these variables are chosen to be $(0.125,0,0,0.75,-1,0,0.1)$. The boundary condition at the outer limits of the domain are reflective.

\begin{figure}
\centering
\includegraphics[width=.5\textwidth]{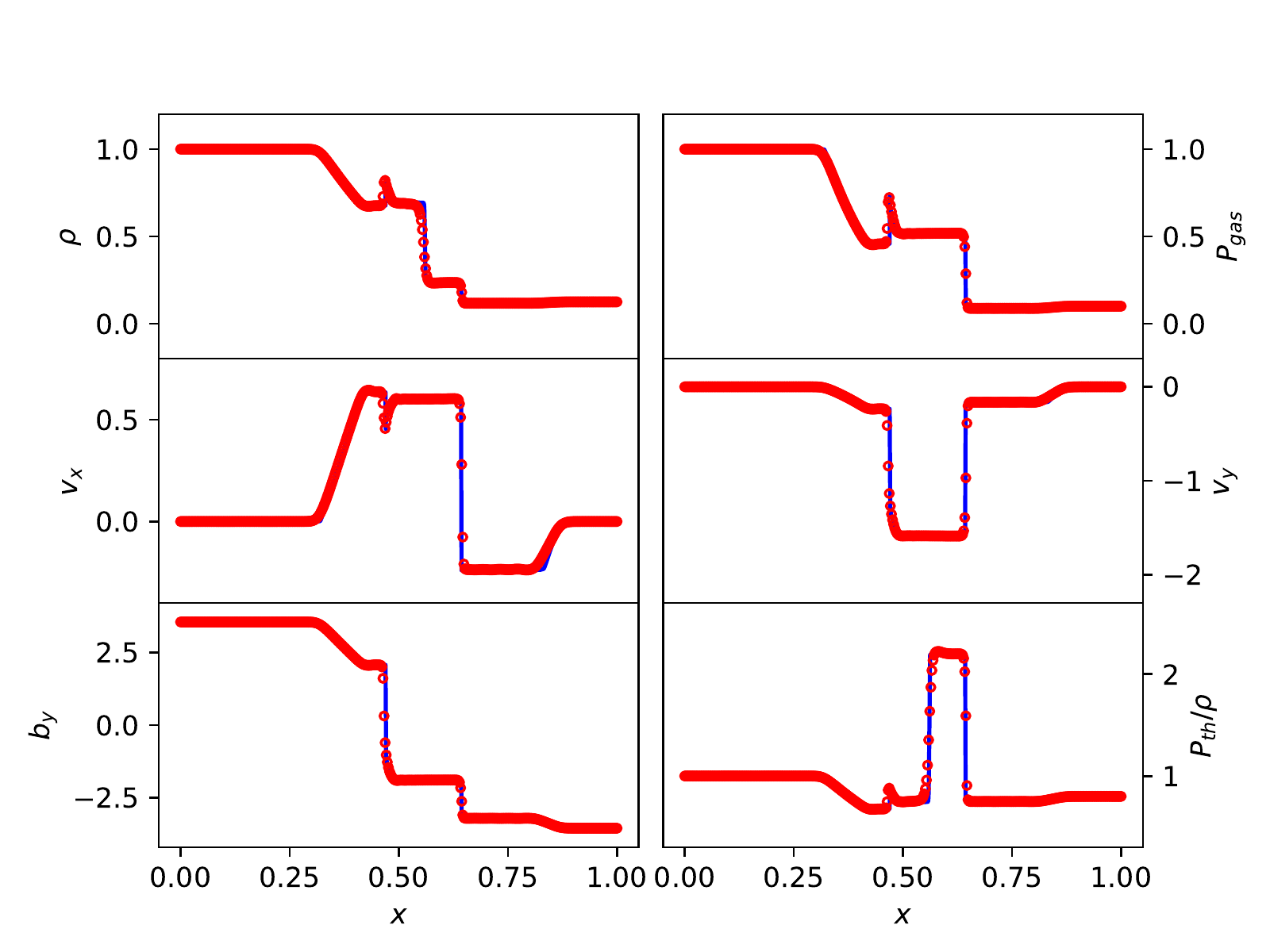}
\caption{Simulation variables as function of the position for the Brio-Wu test at $t=0.1$. Blue lines represent the results for a grid of $10000$ cells while red dots shows the corresponding results for a grid of $400$ cells.}
\label{fig:BWTest}
\end{figure}

Figure \ref{fig:BWTest} shows the final estate of the simulation at $t = 0.1$ using the HLLD solver and the 8-wave method to constrain $\nabla \cdot \mathbf{B}$. The blue lines represent the results for a grid size of $10000$ cells, while the red points shows the results for a grid of $400$ cells.
Our results can be directly compared with fig. 2 of the original work by \cite{BrioWu88}. They can also be compared with the results shown in fig. 13 from the work of \cite{Stone2008}, where the authors implemented this test with the same two resolutions used here into the {\sc athena} code. Our code yields basically the same results.

\subsection*{Kelvin-Helmholtz 2D }

A second test is done in a setup with a shear flow that excites the Kelvin-Helmholtz \citep{Helmholtz,Kelvin} hydro-dynamical instability (and a version modified by magnetic fields).
The test consists of a square domain, $x \in [-0.5,0.5]$ and $y \in [-0.5,0.5]$, filled with two fluids of different densities moving in opposite directions. In the region $|y| < 0.25$ we set $\rho = 2$ and $v_x = 0.5$, while in the region $|y| >= 0.25$ we set $\rho = 1.0$ and $v_x = -0.5$. In the entire domain the pressure has a value of $2.5$ and $\gamma=1.4$. The boundary conditions are periodic everywhere.

To initialise the instabilities a random component to the $v_x$ and $v_y$ variables was added, with a peak-to-peak amplitude of $0.01$.
For the MHD case, the $x$ component of the magnetic field is set to $0.5$.

\begin{figure}
\centering
\includegraphics[width=.4\textwidth]{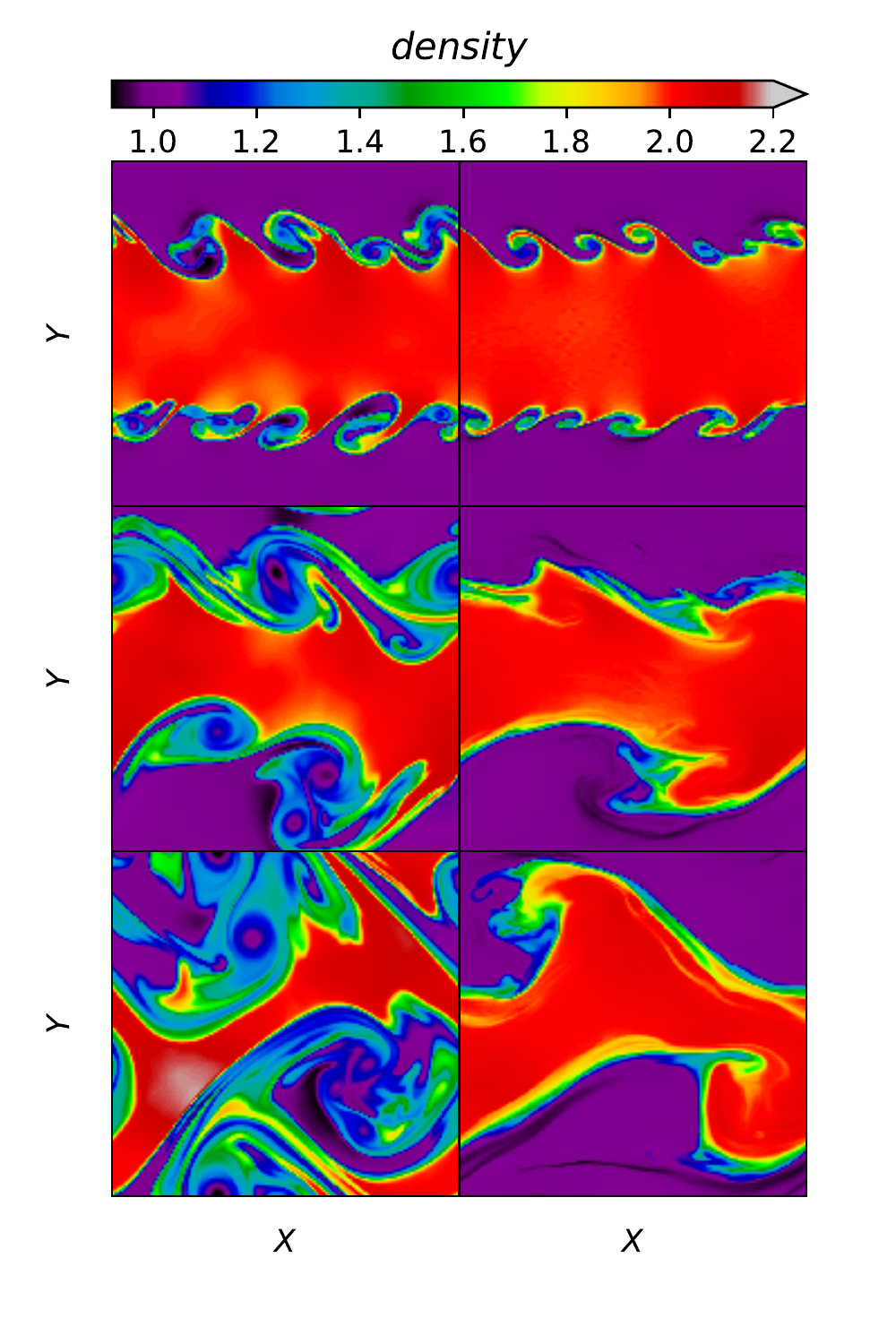}
\caption{Density contours from the Kelvin-Helmholtz test. The left column shows the density for the HD case, while the right column correspond to the MHD case. The rows shows the outputs at $t=1,3$ and $5$ respectively.}
\label{fig:KH}
\end{figure}

Figure \ref{fig:KH} shows the density contours obtained at $t= 1, 3$ and $5$ for a $512\times 512$ mesh. The left column shows the hydrodynamic case while the right column corresponds to the magneto-hydrodynamic case. The test was also run using the HLLD scheme and the 8-wave method.
Our results can be compared with the results presented in the {\sc athena} web page for both cases (HD and MHD) \url{https://www.astro.princeton.edu/~jstone/Athena/tests/kh/kh.html}.

\subsection*{Orszag-Tang 2D}

This vortex system, was originally proposed studied by \citet{1979JFM....90..129O}. Given that the flow develops a complex structure, it is now used extensively as a test to verify the ability of MHD codes to solve turbulence and shocks \citep[see][and references therein]{toth2000,Stone2008}.

The problem consists in a square domain filled with a fluid of density $\rho = 25/(36\pi)$ at a pressure $p = 5/(12\pi)$ and an adiabatic index of $\gamma = 5/3$. The initial velocity and magnetic field components are periodic and set by the functions:
\begin{eqnarray*}
v_\mathrm{x} &=& -\sin(2 \pi y), \\
v_\mathrm{y} &=& \sin(2 \pi x), \\
b_\mathrm{x} &=& -B_0 \sin(2 \pi y), \\
b_\mathrm{y} &=& B_0 \sin(2 \pi x), \\
\end{eqnarray*}
where $B_0 = 1/(4\pi)$. Periodic boundary conditions are set at both sides of the domain.

\begin{figure}
\centering
\includegraphics[width=.4\textwidth]{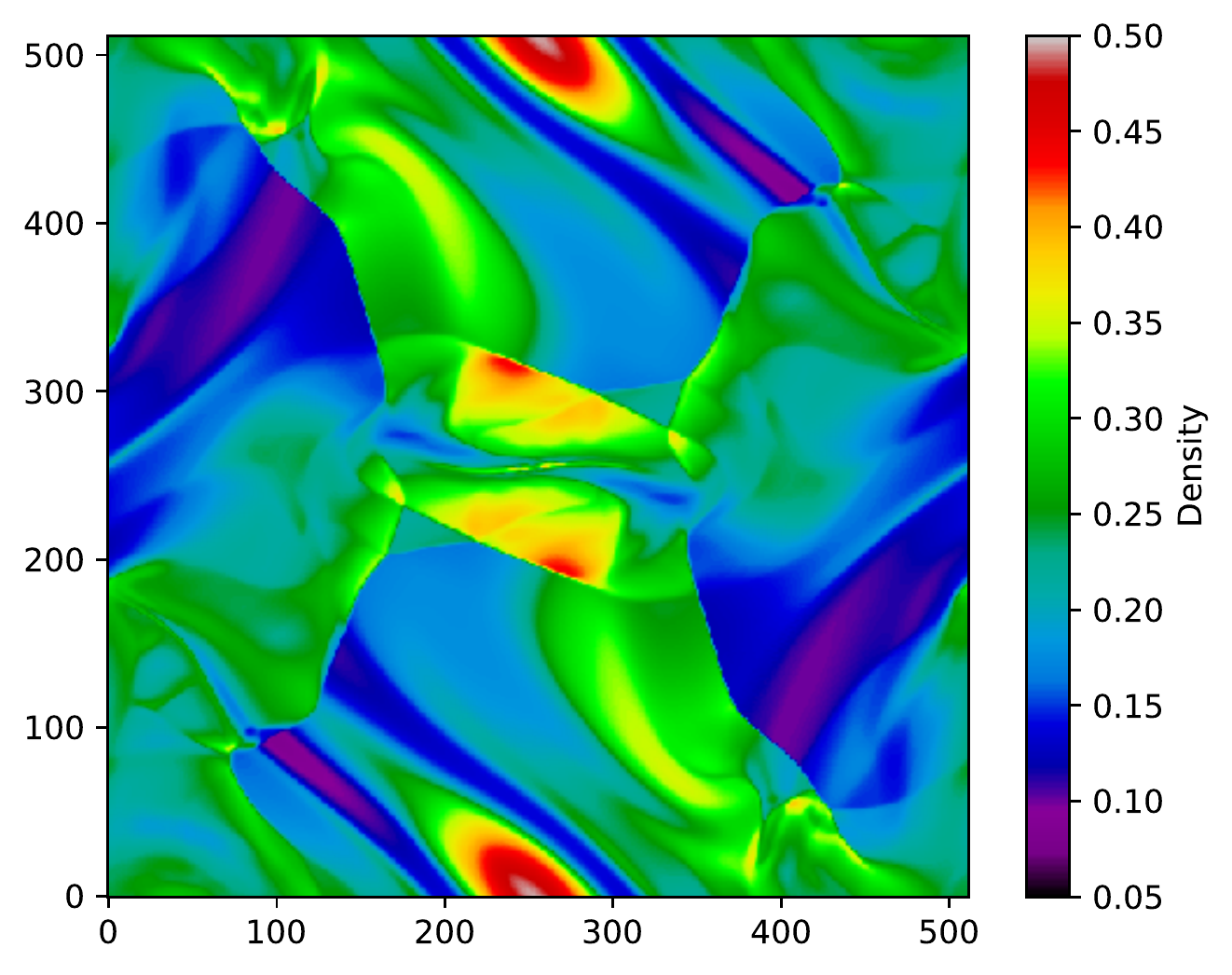}
\caption{Density contours at time $t=0.5$ from the Orszag-Tang test for a 512x512 grid.}
\label{fig:OT}
\end{figure}

Figure \ref{fig:OT} shows the density distribution at time $t = 0.5$ for a 512x512 grid. The result can be compared with those presented in \citet{toth2000,Stone2008}. In this test, to keep the $\nabla\cdot \mathbf{B} = 0$, we used the method of Constrained Transport with the HLLD solver.

\subsection*{Ryu-Jones 3D}

The aim of this test, proposed by \cite{RJ2a}, is to illustrate the ability of the numerical scheme to resolve the presence of fast and slow shocks, as well as rotational discontinuities. It consists in a MHD shock tube with a left state $(\rho,v_x,v_y,v_z,B_x,B_y,Bz,E) = [1.08,1.2,0.01, 0.5,2/\sqrt{4\pi},3.6/\sqrt{4\pi}, 2/\sqrt{4\pi}, 0.95)]$ and a right state $[1.0,0,0,0,2/\sqrt{4\pi},4/\sqrt{4\pi}, 2/\sqrt{4\pi},1)]$. Outflow boundary conditions are used in all the domain and we set $\gamma=5/3$.

\begin{figure}
\centering
\includegraphics[width=.48\textwidth]{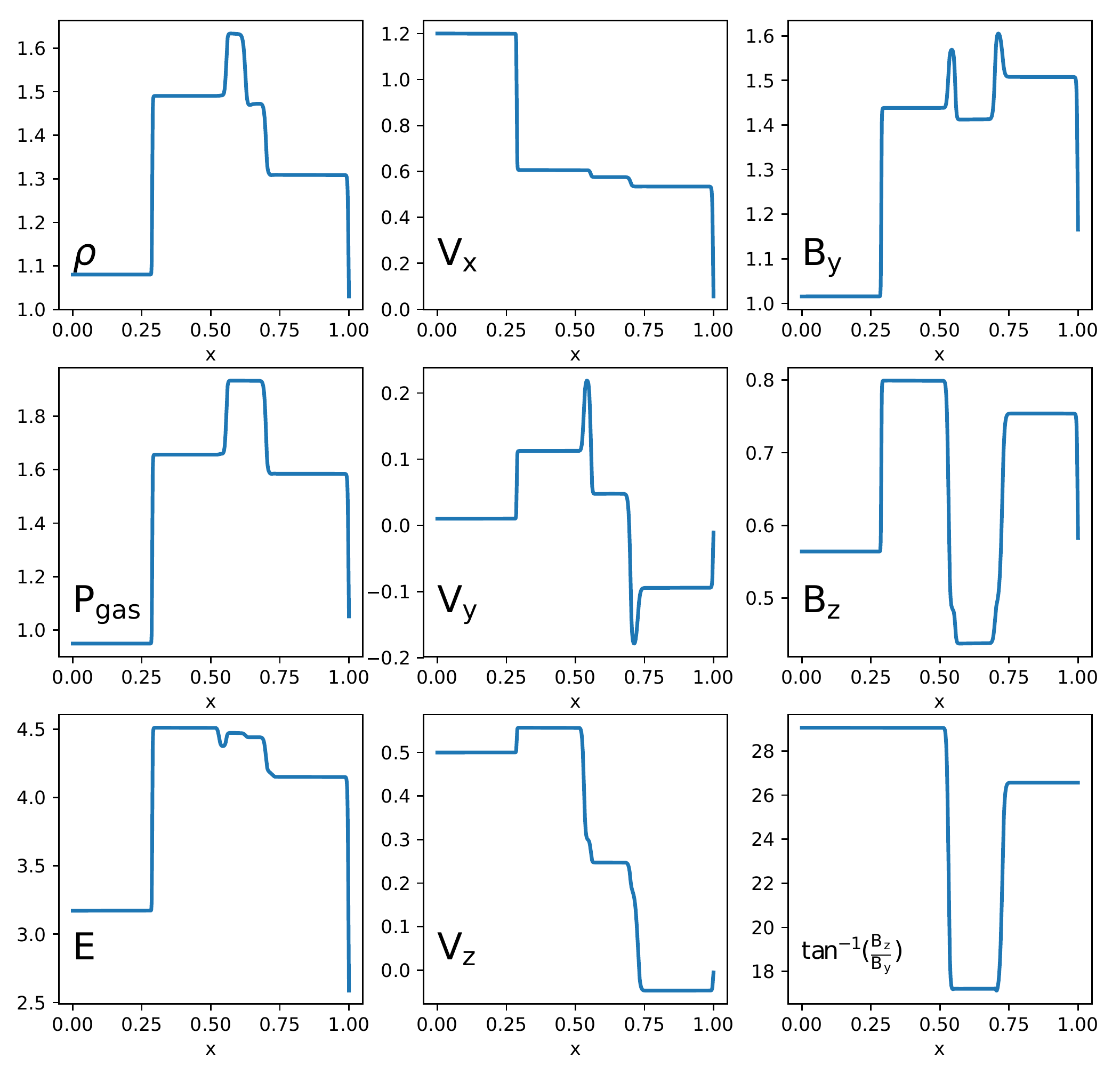}
\caption{Simulation variables as function of the position for the Ryu-Jones 2a test. All the variables are shown for $t= 2.2$ in a grid of $768\times8\times8$ cells.}
\label{fig:rj2a}
\end{figure}

The test was run using the HLLD scheme and the 8-wave method to maintain the $\nabla \cdot \mathbf{B}$ close to zero. The results for $t=2.2$ are shown on Figure \ref{fig:rj2a} for a mesh size of $768\times8\times8$. The size of the mesh was chosen to be able to directly compare our results with those presented in fig. 14 of \citet{Stone2008}.

%%%%%%%%%%%%%%%%%%%%%%%%%%%%%%%%%%%%%%%%%%%%%%%%%%

% Don't change these lines
\bsp	% typesetting comment
\label{lastpage}
\end{document}